\documentclass[lettersize,journal]{IEEEtran}
\usepackage{amsmath}
\usepackage{amssymb}
\usepackage{makecell}
\usepackage{multirow}
\usepackage{booktabs}
\usepackage{tabularx}
\usepackage{amsmath,amsfonts}
\usepackage{float}
\usepackage{algorithmic}
\usepackage{algorithm}
\usepackage{array}
\usepackage[caption=false,font=normalsize,labelfont=sf,textfont=sf]{subfig}
\usepackage{textcomp}
\usepackage{stfloats}
\usepackage{url}
\usepackage{verbatim}
\usepackage{graphicx}
\usepackage{cite}
\begin{document}

\title{Continuous and Noninvasive Measurement of Arterial Pulse Pressure and Pressure Waveform using an Image-free Ultrasound System}

\author{Lirui Xu, Pang Wu, Pan Xia, Fanglin Geng, Peng Wang, Xianxiang Chen, Zhenfeng Li, Lidong Du, Shuping Liu, Li Li, Hongbo Chang and Zhen Fang
\thanks{This work was funded by the National Natural Science Foundation of China (Grant 62071451), National Key Research and Development Project 2020YFC2003703, 2021YFC3002204, 2020YFC1512304, and CAMS Innovation Fund for Medical Sciences(2019-I2M-5-019). \emph{(Corresponding authors: Lidong Du, Shuping Liu, Li Li, Hongbo Chang and Zhen Fang).}

This work involved human subjects or animals in its research. Approval of all ethical and experimental procedures and protocols was granted by the Ethics Committee of Chinese PLA General Hospital under Application No.S2019-318-05.

Lirui Xu, Pang Wu, Pan Xia and Fanglin Geng are with the Aerospace Information Research Institute, Chinese Academy of Sciences,Beijing 100190, China, and also with the School of Electronic, Electrical and Communication Engineering, University of Chinese Academy of Sciences,Beijing 100049, China(e-mail: xulirui19@mails.ucas.ac.cn; wupang17@mails.ucas.ac.cn; xiapan17@mails.ucas.ac.cn; gengfanglin20@mails.ucas.ac.cn).

Peng Wang, Xianxiang Chen, Zhenfeng Li and Lidong Du are with the Aerospace
Information Research Institute, Chinese Academy of Sciences, Beijing 100190, China (e-mail: wangpeng01@aircas.ac.cn; chenxx@aircas.ac.cn; lizhenfeng@aircas.ac.cn; lddu@mail.ie.ac.cn).

Shuping Liu and Li Li are with the Department of Ultrasound, Air Force Medical Center of PLA , Beijing 100037, China(e-mail: kjlsp@163.com; kjlily03@163.com).

Hongbo Chang is with the Department of Neurosurgery, The First Medical Center of Chinese PLA General Hospital, Beijing 100048, China(e-mail: changhongbo@163.com).

Zhen Fang is with the Aerospace Information Research Institute, Chinese Academy of Sciences, Beijing 100190, China, also with the School of Electronic, Electrical and Communication Engineering, University of Chinese Academy of Sciences, Beijing 100049, China, and also with the Personalized Management of Chronic Respiratory Disease, Chinese Academy of Medical Sciences, Beijing 100190, China (e-mail: fangzhen@aircas.ac.cn).
}
}

\markboth{Journal of \LaTeX\ Class Files,~Vol.~14, No.~8, August~2021}%
{Shell \MakeLowercase{\textit{et al.}}: A Sample Article Using IEEEtran.cls for IEEE Journals}


\maketitle

\begin{abstract}
The local beat-to-beat local pulse pressure (PP) and blood pressure waveform of arteries, especially central arteries, are important indicators of the course of cardiovascular diseases (CVDs). Nevertheless, noninvasive measurement of them remains a challenge in the clinic. This work presents a three-element image-free ultrasound system with a low-computational method for real-time measurement of local pulse wave velocity (PWV) and diameter waveforms, enabling real-time and noninvasive continuous PP and blood pressure waveforms measurement without calibration. The developed system has been well-validated in vitro and in vivo. In in vitro cardiovascular phantom experiments, the results demonstrated high accuracy in the measurement of PP (error \textless 3 mmHg) and blood pressure waveform (root-mean-square-errors (RMSE) \textless 2 mmHg, correlation coefficient (r) \textgreater 0.99). In subsequent human carotid experiments, the system was compared with an arterial tonometer, which showed excellent PP accuracy (mean absolute error (MAE) = 3.7 ± 3.4 mmHg) and pressure waveform similarity (RMSE = 3.7 ± 1.6 mmHg, r = 0.98 ± 0.01). Furthermore, comparative experiments with the volume clamp device demonstrated the system's ability to accurately trace blood pressure changes (induced by deep breathing) over a period of one minute, with the MAE of DBP, MAP, and SBP within 5±8 mmHg. The present results demonstrate the accuracy and reliability of the developed system for continuous and noninvasive measurement of arterial PP and blood pressure waveform measurements, with potential applications in the diagnosis and prevention of CVDs.
\end{abstract}


\section{Introduction}
\label{sec:introduction}
\IEEEPARstart{A}{rterial} blood pressure reflects the pathophysiological state of the cardiovascular system, and its related parameters have been well established in the prognostic importance of cardiovascular risk factors\cite{b1, b2}. In the cardiovascular system, periodic pulse pressure oscillates around the mean arterial pressure (MAP), creating diastolic blood pressure (DBP) and systolic blood pressure (SBP). Unlike MAP, which remains constant in the arterial tree\cite{b3}, pulse pressure (PP = SBP - DBP) increases toward the periphery due to the earlier arrival of distal reflected waves\cite{b4, b5}, a phenomenon known as the 'pulse pressure amplification phenomenon.' This indicates that as the distance from the heart increases, the PP in arteries also increases, and the pressure waveform behaves slender morphologically. Pulse pressure amplification is inconsistent with different populations. In elderly and hypertensive patients, increased arterial stiffness led to decreased compliance. It accelerated pulse wave velocity (PWV), which in turn causes the reflected wave and the forward wave of central arterial blood pressure to be superimposed earlier in late systole, which results in elevated central arterial PP and SBP\cite{b4}. It led to a shortened left ventricle diastole period and increased cardiovascular load, leading to severe cardiovascular events\cite{b6}. Therefore, in addition to DBP and SBP, the importance of PP deserves to be noticed, and studies even suggest that PP is a more relevant factor for CVDs\cite{b7, b8}.

Further, the arterial blood pressure waveforms provide extensive information on the cardiovascular condition that is of extraordinary significance for diagnosing CVDs\cite{b9}. Each peak and trough observed in the arterial blood pressure waveform represents a distinct cardiovascular activity\cite{b10}. For instance, the augmentation index (AIx) indicates wave reflection and increased arterial stiffness, helps diagnose early coronary diseases\cite{b11}. The subendocardial viability ratio quantifies myocardial perfusion relative to cardiac workload\cite{b12}. In this way, measuring these minor waveform changes can provide a reliable basis for diagnosing and preventing CVDs.

The use of the brachial artery sphygmomanometer as a reliable and widely accepted technique for measuring blood pressure has been well-established and widely adopted in clinical settings. Although studies show that DBP and MAP do not differ significantly in the arterial tree\cite{b3, b4}, PP and SBP in the brachial artery differ from those in the central artery due to the pulse pressure amplification phenomenon. The parameters of central arterial (e.g., aortic or carotid) blood pressure are more effective predictors of cardiovascular risk than that of peripheral site, because it is a better indicator of the load exerted on the left ventricular and coronary\cite{b13}. For these reasons, measuring blood pressure parameters at the central artery is irreplaceable.

Invasive pressure catheter technology has become the gold standard for blood pressure and waveform measurement due to its accuracy and consistency. However, its invasive nature causes patients pain and is prone to infection\cite{b14}. Volume clamp\cite{b15} and arterial tonometry\cite{b16} are two noninvasive methods to measure arterial blood pressure waveforms. The volume clamp device uses an inflatable finger cuff to compress the finger by controlling the cuff pressure to achieve a state of vascular unloading, in which the sleeve pressure equals the intravascular pressure\cite{b17}. It compresses the vasculature and causes discomfort to patients. In addition, the volume clamp technique is limited to the finger location. Given the importance of central artery pressure parameters, some works attempted to develop models for estimating central arterial blood pressure waveforms from peripheral blood pressure waveforms, such as radial-central artery models\cite{b18, b19} and brachial-central artery models\cite{b20, b21}. However, these methods usually rely on parameter training with large amounts of data. It is hard to exclude the influence of individual cardiovascular differences (e.g., ages, diseases) on the estimation results\cite{b11, b22}. The arterial tonometer reads the transducer pressure values directly by applanating the arterial wall. Due to its anatomical requirements, it can only be used for skeletally supported arteries, mainly the radial artery, followed by the carotid artery. Since it is sensitive to the position, angle, and pressure prevented by the transducer, it requires a well-trained operator to obtain a high-quality signal, which is unsuitable for long-term monitoring\cite{b23}. There is a thick, soft tissue outside the carotid artery, which makes it difficult to determine the optimal pressure applied, and its measurement is highly susceptible to movements such as breathing. Most importantly, the absolute PP and the pressure level measured directly by the arterial tonometer need to be more accurate\cite{b24} and usually require calibration.

Ultrasound technology has been introduced for noninvasive blood pressure waveform estimation due to its ability to penetrate tissue, allowing the detection of the arterial morphology and blood flow velocity. One representative technique involves the use of an empirically derived exponential relationship between arterial pressure and cross-sectional area\cite{b25, b26, b27}. This technology has also been introduced into miniaturized wearable ultrasound arrays\cite{b28}. However, it requires the measurement of SBP and DBP with the aid of an external sphygmomanometer to determine the arterial elasticity coefficient. It lacks continuous calibration capability in the presence of changes in arterial compliance. Another ultrasound-based technique uses local PWV to quantify arterial compliance and thus measure absolute PP and blood pressure waveforms\cite{b29, b30, b31, b32}. Beulen et al. used vertical ultrasound velocimetry to measure blood flow velocity in a fast B-mode ultrasound system to calculate PWV. It implemented complex calculations in offline processing and was only validated in vitro\cite{b29}. Vappou et al. used the pulse wave imaging technique, and the complex local PWV processing was implemented offline too\cite{b30}. These techniques only enable offline or non-real-time blood pressure waveform measurements unless using expensive processing units and complex systems. Seo et al. used two unfocused and wide acoustic beams of ultrasound elements to measure the arterial diameter and blood flow velocity to calculate local PWV to avoid motion artifacts during measurement. A bulky transducer structure is needed in this technique, and the spurious interference from wide acoustic beams poses an accuracy challenge\cite{b31, b32}. Given the clinical value of the blood pressure waveforms, a simple, automated, and easily scalable system is needed be developed for measuring blood pressure waveforms in real-time.

A three-element image-free ultrasound system for real-time and automatic measurement of arterial PP and blood pressure waveform was proposed in this work. Its measurement ability was verified in a series of in vitro and in vivo experiments. The rest of the paper is organized as follows. Section II describes the technique. Section III describes the experimental setup for in vitro experiments on cardiovascular phantom and in vivo experiments on subjects. Section IV shows the results and analysis. Section V discusses this work's significance, limitations, and future scenarios. At last, section VI concludes the paper.

\section{Proposed Technology}
\subsection{Measurement principle}
Arteries are modeled as circular elastic tubes, while the periodic blood flow generated by left ventricular contraction is modeled as pulsatile flow. Pulsatile flow travels through the elastic artery as the cross-sectional area of the artery distends in response to changes in intra-arterial pressure. The artery compliance (AC = dA/dP) reflects the magnitude of this distension. According to the Bramwell-Hill equation, the relationship between PWV and compliance can be expressed as\cite{b33}:
\begin{equation}
\label{deqn_1}
PWV = \sqrt{\frac{A}{\rho} \frac{dP}{dA}},
\end{equation}
where \textit{P} is the arterial blood pressure, \textit{A} is the arterial cross-sectional area, and $\rho$ is the blood density ($\rho$ usually takes 1060 kg/m$^3$). The arterial blood pressure \textit{P(t)} is expressed by PWV and the arterial cross-sectional area \textit{A(t)}\cite{b29,  b30}.
\begin{equation}
\label{deqn_2}
P(t) - P_0 = \rho PWV^2\ln (\frac{A(t)}{A_0}),
\end{equation}
where $P_0$ and $A_0$ are the blood pressure and the arterial cross-sectional area at the exact moment, for example, the diastolic blood pressure DBP and the end-diastolic arterial cross-sectional area currently in $A_d$. The large artery filled with blood is approximately circular for a healthy individual, the area waveform \textit{A(t)} is calculated from the diameter waveform \textit{D(t)}.
It means that measuring the arterial diameter of the artery in real-time, combined with the local PWV at the measurement site, gives a real-time PP and pressure waveform, PP is expressed as:
\begin{equation}
\label{deqn_3}
PP = 2\rho PWV^2\ln (\frac{D_s}{D_d}).
\end{equation}
where $D_s$ and $D_d$ are the arterial end-systolic and end-diastolic diameters, respectively. Further, since DBP varies minimally in the arterial tree\cite{b4, b34}, the brachial DBP can be measured using a sphygmomanometer to convert the PP waveform at the measured arterial location into an absolute blood pressure waveform.

\begin{figure*}[!t]
\centering
\includegraphics[width=18.1cm]{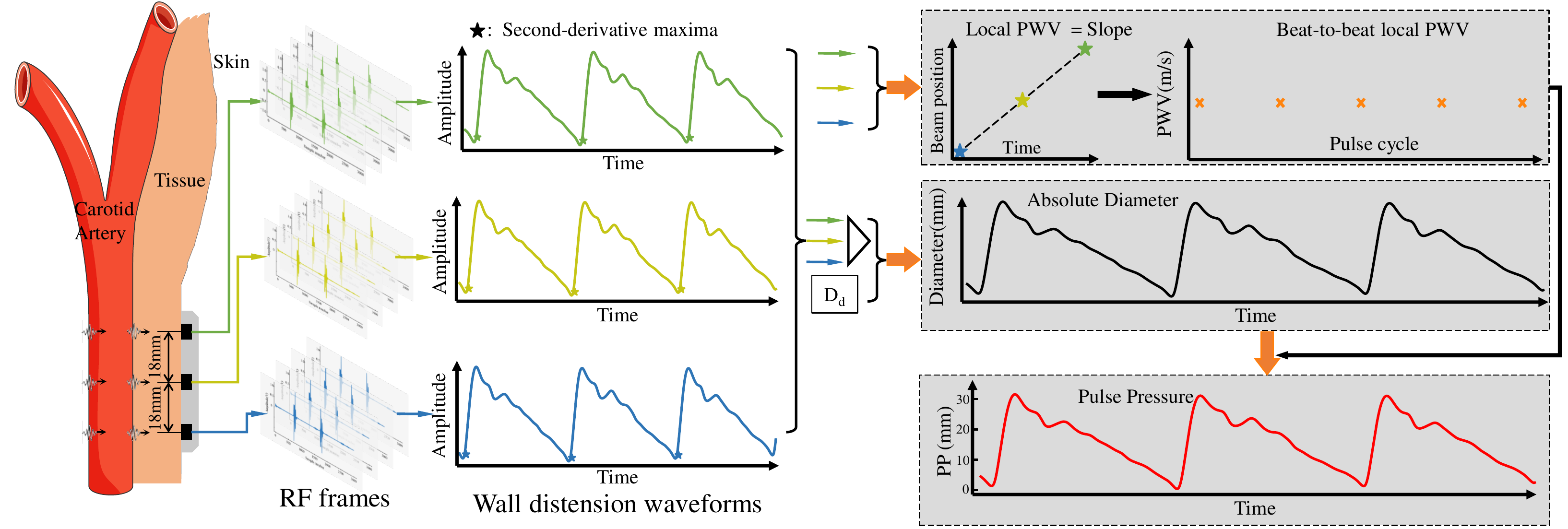}
\caption{ Measurement principle of PP and blood pressure waveform.
\label{fig_1}}
\end{figure*} 
\unskip

In this work, a three-element ultrasound probe measured local PWV and arterial diameter, and PP and blood pressure waveforms were calculated. Each array element has a diameter of 3 mm, a center frequency of 5 MHz, and a beam half angle of only 1.7°. The beam half-angle means that when the ultrasound wave exceeds this angle in the axial direction, the acoustic intensity decays by more than half, indicating beam directivity. Because the artery is circular, the half angle of 1.7° makes the ultrasonic probe receive a strong echo only when it is perpendicular to the artery and aligned with the artery diameter. Therefore, all the echoes recorded by the probe are due to the echo when the probe is aligned with the artery diameter. The distance between each adjacent ultrasound element is 18mm. When the pressure pulse propagates in the artery, causing the proximal artery to distend earlier than the distal artery, the transmission delay between the three locations of the distension waveform (measured as the time difference between the points of the second-order derivative peaks) is used to calculate the local PWV and to assess arterial compliance. While a two-element probe can also measure local PWV, using multiple channels can reduce random errors to a certain extent and improve measurement accuracy. Furthermore, since PWV is not constant throughout the artery, combining the local PWV measured by the three array elements with the diameter of the intermediate element can better reflect the compliance of the intermediate position. The arterial distension waveform is combined with the absolute diameter at a given moment to obtain a real-time arterial diameter waveform. The obtained local PWV and diameter waveforms are substituted into \eqref{deqn_2} and \eqref{deqn_3} to calculate PP and blood pressure waveforms. It is worth noting that arterial compliance does not change significantly over a short time, so local PWV is only assessed at the beginning 10 seconds and is measured at regular intervals in subsequent measurements (i.e., every 3 minutes). Fig. 1 summarizes the PP waveform measurement principle in this work.

\subsection{Measurement Hardware}
The custom integrated ultrasound transceiver hardware system interfaced with the ultrasound probe for excitation, pre-processing, acquisition, and transmission of the ultrasound transducer signal. The hardware architecture is illustrated in Fig. 2(a). Operating in pulse/echo mode, the system received echo frames after ultrasound excitation with a configured frame rate or pulse repetition frequency (PRF) of 2000 Hz, controlled by an Altera Stratix IV FPGA (Intel, Santa Clara, CA, USA). The high-voltage pulse transmitter chip TX7316 (Texas Instruments, Dallas, TX, USA) was switched to pulse transmitting mode (TX), emitting a 60 V square wave with a width of 100 ns to excite the ultrasound transducer to emit an ultrasound wave. After 4 us, TX7316 switched to receiving mode (RX). The analog echo signal (40us) was pre-processed and digitized using the ultrasound-specific analog front-end AFE5818 (Texas Instruments, Dallas, TX, USA). The echoes were passed through a high-pass filter (HPF, 200 KHz) and a low-pass filter (LPF, 10 MHz) in the AFE5818, amplified by a low-noise amplifier (LNA, 24 dB), and a programmable gain amplifier (PGA, 24 dB). To ensure synchronized ultrasound excitation and echo reception processing, multiple channels were used to avoid time delays between channels. The processed analog signals were digitized by the AFE5818's internally integrated 80Msps ADC and transferred to the FPGA. The echo signals were then transferred to the PC via a super-speed USB 3.0 chip (CYUSB3014, Infineon, Neubiberg,Germany) and processed in real-time using the measurement software. The electronics hardware board is shown in Fig. 2(b).

\begin{figure*}[!t]
\centering
\includegraphics[width=18.1cm]{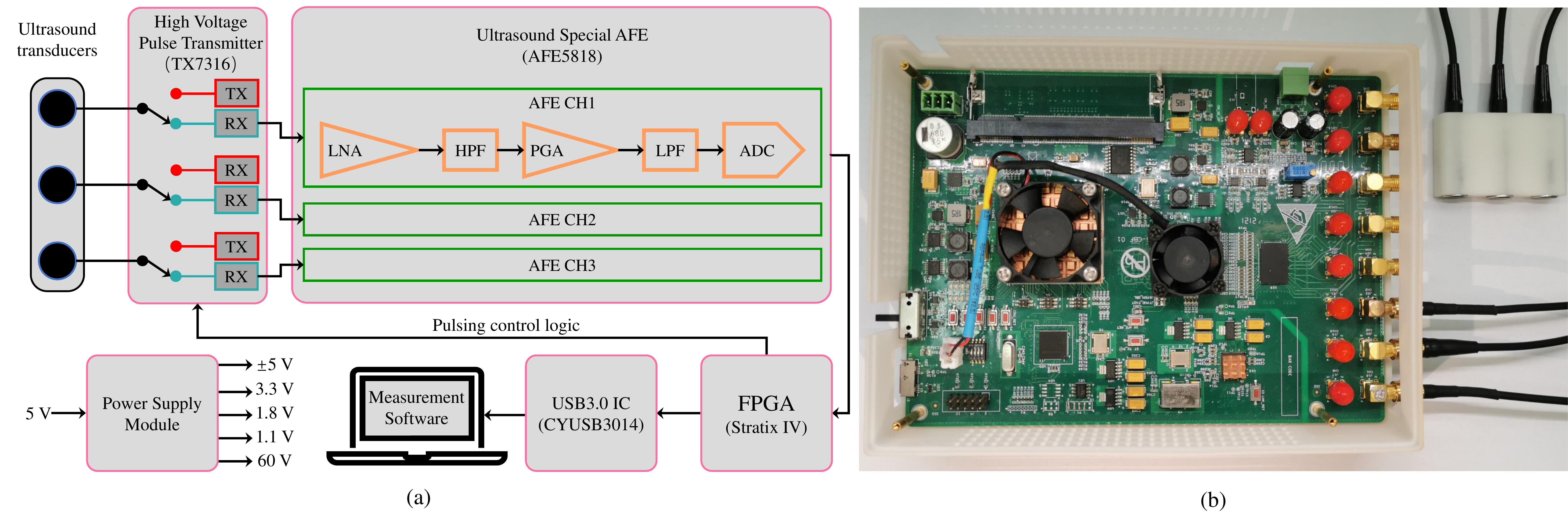}
\caption{(a) Ultrasound hardware system architecture. (b) Electronics hardware board
\label{fig_2}}
\end{figure*} 
\unskip

\subsection{Measurement Software}
A measurement software with a graphical user interface (GUI, Fig. 3) was developed for online signal processing, blood pressure calculation, and data display. For ultrasound probes without images, the inability to correct the position of the probe with the help of images poses a challenge for easy use. To address this limitation, the software identifies the arterial wall using the automatic cross-correlation-based arterial wall identification technique developed by Sahani et al.\cite{b35} to guide the operator to place the probe in the correct position. Once the software identified that all three ultrasound elements received echoes from the anterior and posterior arterial walls and that the signal-to-noise ratio (SNR) was above 15 dB, local PWV and blood pressure assessments were performed. SNR = 20 × log (mean absolute amplitude of wall echo / mean absolute amplitude of the noise above 5mm depth).

\begin{figure}[!t]
\centering
\includegraphics[width=3.5in]{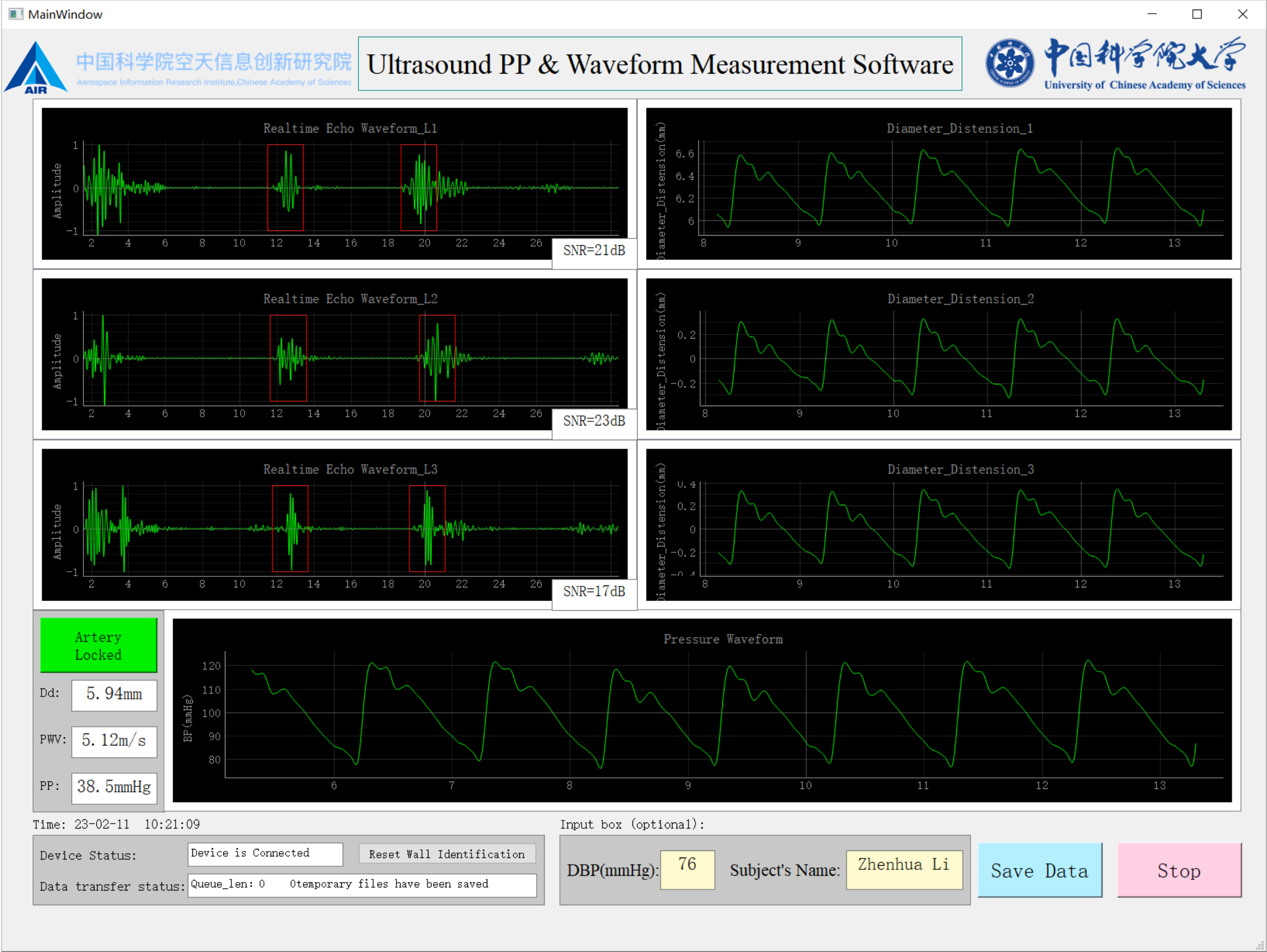}
\caption{ A screenshot of the software GUI.
\label{fig_3}}
\end{figure} 
\unskip

The blood pressure waveform calculation consists of the local PWV online evaluation phase and the real-time waveform tracing phase. Three channels acquire the raw ultrasound signal at a high PRF of 2000 Hz. Such a frame rate is necessary for the local PWV calculation based on time differences between channels, which are typically a few milliseconds. For blood pressure waveforms, a PRF of 200 Hz or more for a single channel is sufficient to acquire sufficient waveform detail.
\begin{figure}[!t]
\centering
\includegraphics[width=3.5in]{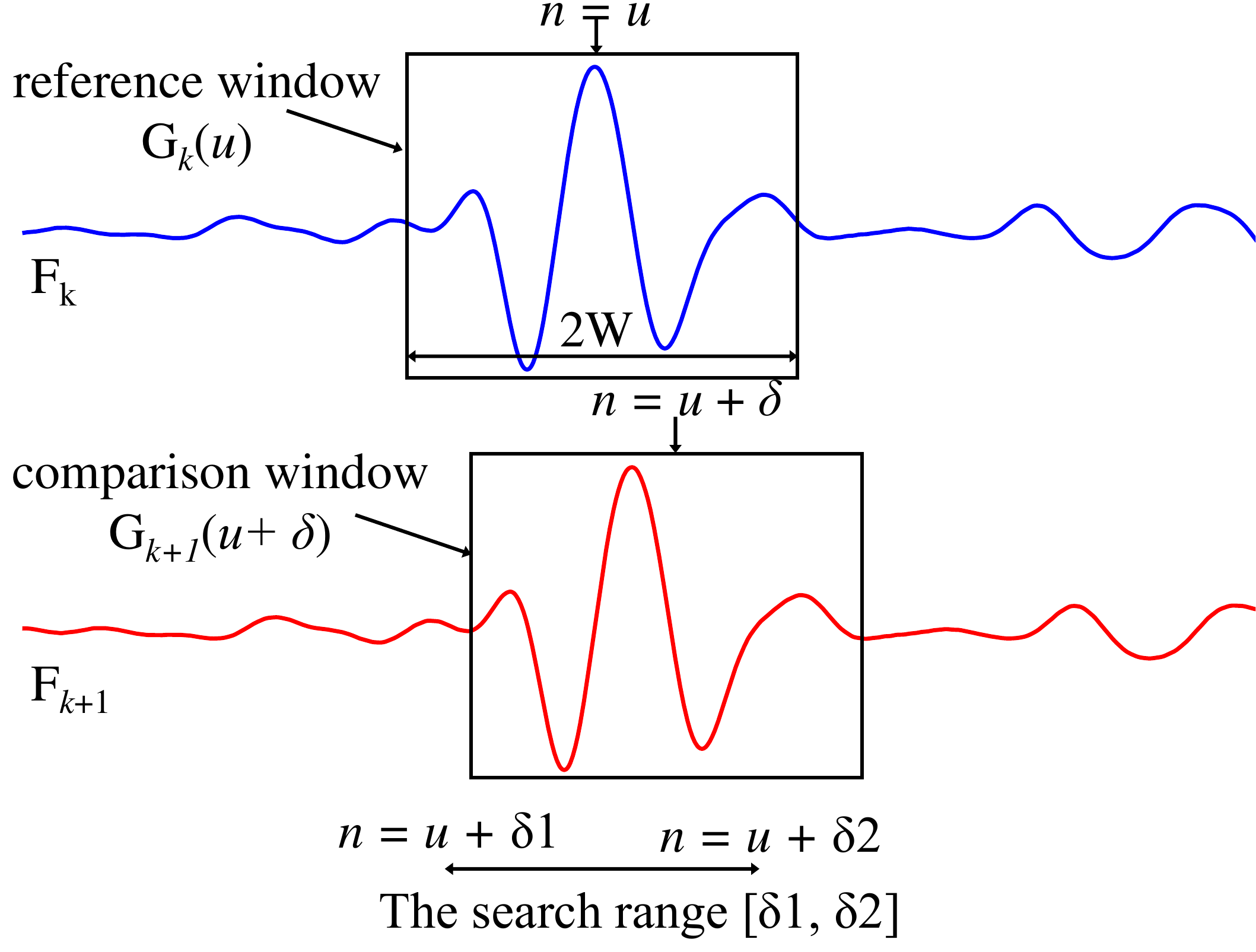}
\caption{ Principle of the cross-correlation algorithm.
\label{fig_4}}
\end{figure} 
\unskip

The evaluation of the local PWV relies on estimating the arterial wall distensions. The amount of arterial wall shift between two adjacent echo frames is calculated using a cross correlation algorithm\cite{b36}. The fundamental principle of the cross-correlation algorithm is to find the maximum similarity between two signals by shifting one signal relative to the other and calculating the correlation coefficient at each shift. For the \textit{k}th echo frame $F_\textit{k}$, where the approximate area of a single arterial wall (anterior or posterior) located is denoted as reference window $G_\textit{k}$,
\begin{equation}
\label{deqn_4}
G_k(u) = F_k(u-W+n),\ \ \ \ (0<n\le2W).
\end{equation}

$G_\textit{k}$ is an echo segment with a window length of 2W centered at \textit{u}. \textit{u} is usually chosen as the peak or envelope peak of the region of the arterial wall echo located. Assuming that the shift of the arterial wall between frames \textit{k} and \textit{k}+1 is \textit{s}, the goal of the cross-correlation shift estimation is to estimate \textit{s}. By sliding the echo window of the arterial wall at frame \textit{k}+1 with a sliding amount of $\delta$ samples, the comparison window is constructed as follows:
\begin{equation}
\label{deqn_5}
\begin{split}
    G_{k+1}(u+\delta) = F_{k+1}((u+\delta)-W+n), \\ (\delta_1\le \delta \le \delta_2,0<n\le2W).
\end{split}
\end{equation}
In order to estimate the correct arterial wall shift \textit{s}, the cross correlation \textit{C}($\delta$) is calculated as:
\begin{equation}
\label{deqn_6}
\begin{split}
    C(\delta) = \sum_{n=u-W}^{u+W} G_{k}(\delta) G_{k+1}(u+\delta), \ \  \  (\delta_1\le \delta \le \delta_2).
\end{split}
\end{equation}
The $\delta$ such that \textit{C}($\delta$) takes the maximum value is the estimate of \textit{s}, while [$\delta$1, $\delta$2] is the search range. It is worth noting that [$\delta$1, $\delta$2] must cover the arterial wall shift in the physiological range, and the probe during the measurement should also be considered. Fig. 4 illustrates the principle of the cross-correlation algorithm.

The high frame rate data and the large number of cross-correlation calculations involved in the above process make it impossible to implement real-time local PWV calculations on a conventional CPU. In order to enable real-time blood pressure acquisition, the computation of high frame rate distension needs to be reduced. Three parameters in the above-presented cross-correlation algorithm significantly affect the computational effort: frame rate, window length 2W, and cross-correlation search range [$\delta$1, $\delta$2]. To reduce computational effort and achieve real-time evaluation of blood pressure, this work employs several strategies to optimize the algorithm:

1) The high-resolution (PRF=2000Hz) distension waveform evaluation is implemented only near the maximum point of the second-order derivative, and the rest of the region is implemented at a low frame rate (PRF = 100 Hz).

2) The window length covers only 2 cycles of echoes near the highest peak of the arterial wall echo and no longer the entire envelope of the echo segment. For a 5 MHz echo at a sampling rate of 80 MHz, the window length is 32 points, and after interpolation to 1.2 GHz, the window length is 480 points.

3) The approximate shift between two frames was first located using the peak to narrow the search [$\delta$1, $\delta$2].

Since the local PWV depends on the estimation of the time difference between channels, for the time difference between different channels is usually a few milliseconds, a high PRF echo is necessary to ensure the accuracy of the local PWV. In this work, the PRF is configured to 2000 Hz. The calculation of the time difference of the distension waveform between different arterial locations is based on the second-order derivative maximum point, so the evaluation of a high frame rate needs to be implemented only near the second-order derivative maximum point\cite{b37}. Therefore, the local PWV calculation was divided into two stages: a low frame rate pre-processing stage and a high frame rate calculation stage. During the first stage, raw data at a rate of 2000 Hz was stored in memory, and in another thread, the data was simultaneously extracted at a rate of 100 Hz to evaluate the low-resolution distension waveform, which was used to determine the approximate range for high frame rate evaluation. The start and end frames of the range were recorded, corresponding to the first point and last eight points of the 100 Hz minimum, which covered an 80 ms range, or 160 frames of the high frame rate. During the second stage, the original frames in memory were evaluated for high frame rate using the recorded range from the first stage. The raw echo frames were unsampled to 1.2 GHz using cubic spline interpolation, and the high-resolution distension waveform was calculated using the cross-correlation method described above. Further, since the evaluation of time difference for each cardiac cycle is independent of each other, the high frame rate calculation for different cardiac cycles is done in separate threads to ensure real-time calculation of local PWV, thereby significantly reducing computation time.

Further, in previous studies, the window length for estimating arterial wall shift typically covered the entire echo envelope, i.e., 1 mm to 1.5 mm of data before and after the peak of the arterial wall, and for data interpolated to 1.2 GHz (the descriptions that follow are all default echo data interpolated to 1.2 GHz), corresponding to a window length of 1560 to 2340 points. Such a window covering the entire arterial wall echo segment provides higher robustness in the wall shift estimation and achieves higher accuracy even when the signal quality deteriorates, but also brings a greater computational consumption. When the window is reduced, significant savings in computational computing resources can be achieved. In previous studies, it was even possible to represent the arterial wall motion by tracking the main peak point (i.e., 2*W=1) when the signal-to-noise ratio of the echo signal was high\cite{ b39}, but the echoes were sometimes distorted, leading to incorrect shift estimation\cite{b39}. In this work, the window length was reduced to 2 echo periods, i.e., a window length of 480 points, significantly reducing the computational cost of a single calculation. Since the SNR of the echo signal is higher than 15 dB, and the high frame rate evaluation is performed on only a small fraction of the signal segments, the cumulative effect of multiple cross-correlation shift estimation errors for shorter window lengths is insignificant.

Finally, during the interdependent wall tracking, the windows [$\delta$1, $\delta$2] of the sliding window cyclic interdependent calculation in (6) must cover the maximum shift in both positive and negative directions and consider the probe movement. An extensive search range leads to most of the calculations being redundant since the actual shift of the arterial wall is not previously known. In this work, the approximate shift of the pipe wall between the two frames is first determined using a peak finding method to reduce the search range to achieve a minor computational effort. It is implemented by first searching for the maximum point within the region of interest of the anterior and posterior walls of the artery in the initial frame and subsequent echo frames, continuously tracking the position of this point of the peak and the shift of the point of the peak between frames $\delta$\textit{p}. If it is within a reasonable physiological shift range, the search range is [$\delta$\textit{p}-3, $\delta$\textit{p}-3], centered on $\delta$\textit{p}. Otherwise, it means that the probe has been displaced or the echo waveform morphology has changed, causing the peak point to be mismarked, at which time the traditional method is used with a more extensive search range.

The high-resolution distension waveform calculated using the above method is then passed through a 4-order low-pass filter with a cutoff frequency of 16 Hz. The filter applies a linear digital filter twice, once forward and once backward, to achieve zero phase shift, ensuring the time points avoid time-shifted before and after the filtering. After filtering, the distension waveform is up sampled to 10 KHz. Finally, the local PWV is calculated by linear regression of the three beam positions and time reference points. This process lasts 10 seconds, and the local PWV used to calculate the blood pressure is the mean of the 10 seconds of data.

After obtaining the local PWV, echo frames are captured at 200 Hz using the element with the highest signal-to-noise ratio in the arterial echo segment. Then the arterial diameter waveform is calculated. The calculation of the arterial diameter waveform includes two steps, the diameter distension waveform and the absolute diameter calculation at a given moment. The arterial diameter waveform \textit{D(t)} was estimated from the echo frames using the cross-correlation algorithm described above and the envelope threshold-based end-diastolic diameter measurement ($D_0$) algorithm\cite{b40}. The local PWV and $D_0$ are substituted into (2) to convert \textit{D(t)} to a blood pressure waveform in real-time.

\section{Experiment Setups}
\subsection{Phantom Experiment and Setups}
The gold standard for measuring blood pressure and pressure waveforms involves invasive pressure catheters, which can cause infections in subjects, making it challenging to implement invasive gold standards in humans. The cardiovascular phantom aims to validate the proposed technology using invasive technologies under controlled conditions.
\begin{figure}[htbp]
\centering
\includegraphics[width=3.5in]{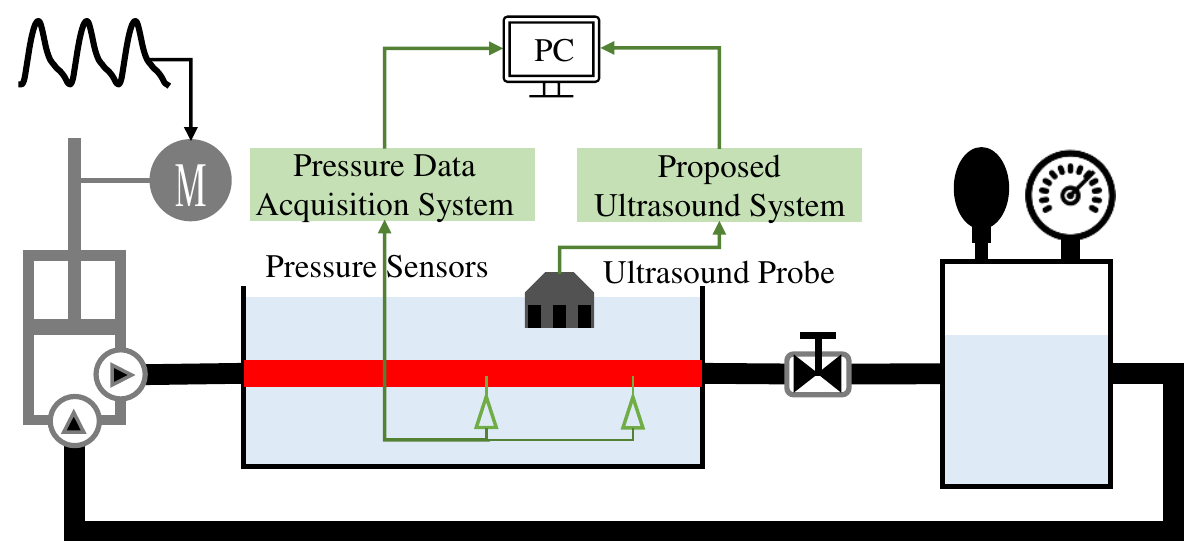}
\caption{Cardiovascular phantom setup.
\label{fig_5}}
\end{figure} 
\unskip

The cardiovascular phantom consists of a pulsatile pump, a silicone vascular model, a half-full sealed tank, two pressure sensors, connecting tubing and valves. A schematic diagram of the cardiovascular phantom setup is illustrated in Fig. 5. The pulsation pump is designed in-house and consists of a high-precision servo motor and a piston pump. The servo motor is connected to a computer, under the control of which the servo motor pushes the piston to erupt fluid in a set trajectory to simulate left ventricular ejection. A soft silicone tube with an inner diameter of 8.2 mm, a wall thickness of 0.4 mm, and a length of 1 m is used to simulate an artery that is completely submerged in water. One end of the silicone tube was connected to the pump outlet via another soft silicone tube, and the other end was connected to a half-full sealed tank via a valve. The valve simulates the resistance to flow at and at the end of a small artery while generating a reflected wave that makes the pressure waveform in the silicone tube closer to the actual arterial pressure waveform. Adjusting the position and degree of opening of the valve can alter the shape of the falling edge of the waveform. The parameters related to the blood pressure waveform (heart rate, blood pressure, waveform shape) can be changed by changing the pump output trajectory, valve position, and valve opening degree. A half-full sealed tank stores the water, while a hand pump can adjust the base pressure inside to generate different DBP. Two medical invasive pressure sensors (Fluid-filled catheter, PT-1, Lontek  Technology, Shenzhen, China) insert into the silicone tubing through needles, located 30 cm (P1) and 20 cm (P2) from the arterial outlet, to measure the pressure waveform and local PWV. The two pressure sensors were digitized synchronously at a sampling frequency of 20 KHz by a high-speed data acquisition card (USB3131A, ART Technology, Beijing, China) and transferred to the data acquisition software of the computer. The ultrasound probe is located between two pressure sensors on the silicone tube, 1.5 cm away from them, with a distance of 3 cm from pressure sensor P1.

\subsection{PP and waveform accuracy compared to tonometer}

In vivo experiments were performed to validate the accuracy of the proposed system for in vivo measurement of PP and blood pressure waveforms. The arterial tonometer can noninvasively measure local PP and blood pressure waveforms at specific locations. The arterial tonometry device SphygmoCor EM3 (AtCor Medical, Sydney, Australia) was used as a comparison standard device. 20 healthy subjects (15 males and 5 female,) aged from 21 to 34 years(25.4 ± 3.2 years) and body mass index (BMI) from 18.3 to 28.9  kg/m$^2$(22.5 ± 2.4 kg/m$^2$) participated in this study. The measurement was taken at the carotid artery, which is the central artery location that is relatively easy to measure with our ultrasound probe, and the only central artery location that can be directly measured with the arterial tonometer. The test was performed in a room with a temperature of 22-24°C. Before the experiment, subjects rested supine for at least 10 minutes to ensure hemodynamic stability. The test was performed with the subject lying flat on the bed with the head slightly tilted back to fully expose the carotid artery.

Before the arterial tonometer measurement, the left arm blood pressure was measured using the SunTech Oscar2 sphygmomanometer (SunTech Medical, Raleigh, North Carolina, USA), and the DBP and MAP were calculated using the AccuWin$^{TM}$ Pro4 software provided by SunTech. The recorded DBP and MAP were entered into SphygmoCor Software (Version: 8) to calibrate the absolute values of the arterial tonometer. Immediately afterward, SphygmoCor EM3 tonometer pen was measured at the subject's carotid artery to measure PP and pressure waveform. The operator adjusted the position of the tonometer pen and recorded the waveform when the stable waveform was captured for at least 11 seconds. To ensure the signal quality, the data is only valid when the 'Operator Index' displayed by the software is higher than 90, otherwise the data was recaptured. For each subject, the software automatically generated average PP and blood pressure waveforms using 10 seconds of data. Immediately following the ultrasound, the operator placed the ultrasound probe in the same position and made fine adjustments until a strong arterial wall echo appeared on the screen. When software identified that all three elements received echoes from the anterior and posterior arterial walls and that the SNR was above 15 dB, the software automatically begins to measure and display the local PWV, PP, and blood pressure waveforms. The original echo data of 11 seconds were recorded for comparative analysis.

\subsection{Tracing blood pressure changes compared to volume clamp device}
In this part, to verify the ability of the proposed ultrasound system to trace blood pressure changes, ultrasound device and volume clamp device Nexfin (Edwards Lifesciences, Irvine, CA, USA) were used to simultaneously measure blood pressure changes in subjects during deep breathing movements. 7 healthy subjects (6 males and 1 female, age = 26.3 ± 3.7 years, BMI = 23.1 ± 3.1 kg/m$^2$) participated in this study. The requirements for the subjects, the measurement environment, and the measurement position were consistent with those described in the previous section. 

After the subject wore a Nexfin finger cuff on the middle finger of the left hand, when the Nexfin device completed its automatic calibration and output a stable blood pressure waveform, the operator used the ultrasound probe to perform measurements in the left carotid artery. The software recognized the arterial wall and began the measurement when the operator placed the probe at an optimized place. First, the subject breathed normally for 15 s, followed by deep breaths (each breath cycle was approximately 10 s) for 45 s, the total test duration was one minute. The data collected from two systems (ultrasound and Nexfin) were synchronized using the local time of the data acquisition computer. When the measurement was completed, the Nexfin and ultrasound data were saved for statistical analysis. The Nexfin was measured at the finger artery and transformed finger pressure to blood pressure by its internal transfer function. There is a significant difference between brachial SBP and carotid SBP, so Nexfin brachial SBP was transformed to carotid SBP. The transform method is described in section IV-C. 

\subsection{Statistical analysis}
In the analysis of the results, all mean values were quantified using mean and standard deviation (SD), i.e., Mean±SD. Bland-Altman analysis was performed to investigate the measurement differences, and degree of variation between the measured values of the two devices, also quantified using root mean square error (RMSE). The DBP, MAP, and SBP comparisons with the standard device were quantified using mean absolute error (MAE) and SD, i.e., MAE ± SD. A 4-order Butterworth low-pass filter filtered all blood pressure waveforms with a cutoff frequency of 25 Hz to remove high-frequency interference. For assessing waveform accuracy, RMSE and correlation coefficient (r) were used to determine the similarity of the waveforms measured by the two devices.

\section{Result}
This section presents the main experimental results. IV-A demonstrates the performance in the cardiovascular phantom. IV-B demonstrates the accuracy of PP and blood pressure waveforms compared to the arterial tonometer. IV-C demonstrates the accuracy of tracing blood pressure changes compared to the volume clamp device in the presence of deep-breathing changes in blood pressure.
\subsection{In vitro Performance Validation}
Fig. 6 shows a comparison of the pressure waveforms measured using the ultrasound method (red line) and the pressure sensor P1 (black line) for a pulsation pump output according to a specific waveform with a frequency of 1 Hz (noted as Trial 1). Since only the absolute PP was measured in this work, the minimum pressure measured by ultrasound in the first cycle was aligned with the minimum pressure of the pressure sensor. Figure (b) shows the difference between the peak pressure measured by the two methods in the 10th cycle, with a difference of 2.73 mmHg. The average PP measurement error over the ten cycles was 2.83 ± 0.04 mmHg.The RMSE between the pressure waveform measured by ultrasound and the reference waveform was 1.5 mmHg, and the correlation coefficient of the waveform is r = 0.995. 

\begin{figure*}[!t]
\centering
\includegraphics[width=18cm]{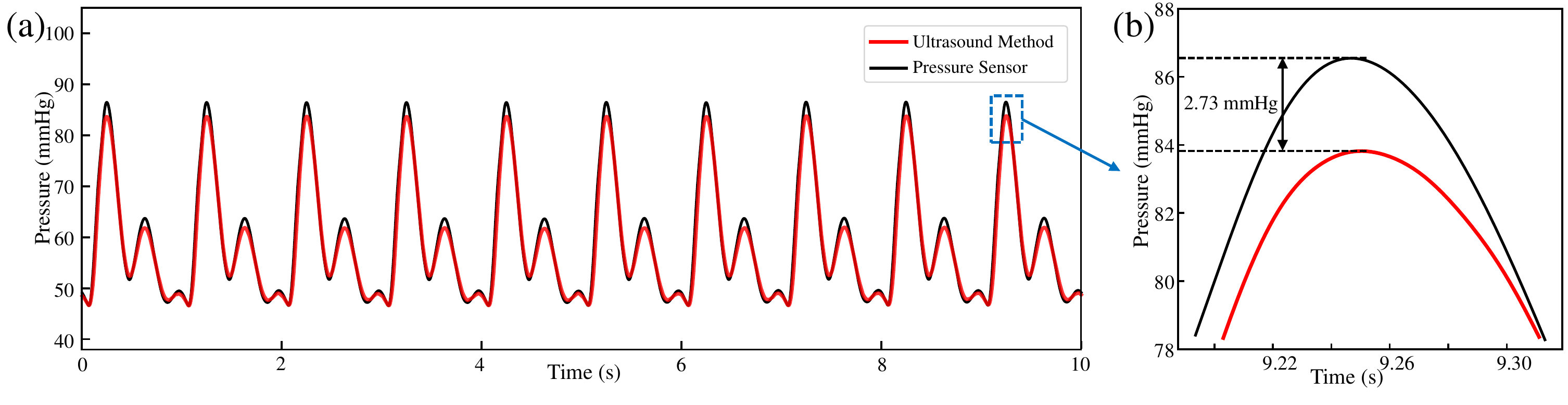}
\caption{ Pressure waveforms measured using the ultrasound method and the pressure sensor.
\label{fig_6}}
\end{figure*} 
\unskip

To further validate the technology’s accuracy, the phantom system settings were changed to output a typical carotid waveform at 60 Hz with base pressures of 63 mmHg (Trial 2) and 88 mmHg (Trial 3), respectively. Fig. 7 demonstrates the measurement results of the ultrasound method. Again, the minimum value of the first cycle was aligned with the minimum value of the pressure sensor. The PP errors for Experiment 2 and Experiment 3 were 0.52 ± 0.05 mmHg and 2.25 ± 0.06 mmHg, respectively, and the RMSEs were 0.91 mmHg and 0.90 mmHg, respectively, compared with the pressure sensor measurement waveforms. The correlation of the waveforms for both sets of experiments was 0.998.  Table 1 shows all the measurement results of the three groups of experiments, and the comparison between the PWV measurement value (PWV$_U$) and the reference value (PWV$_R$) of the pressure sensor.

\begin{figure*}[!t]
\centering
\includegraphics[width=18.1cm]{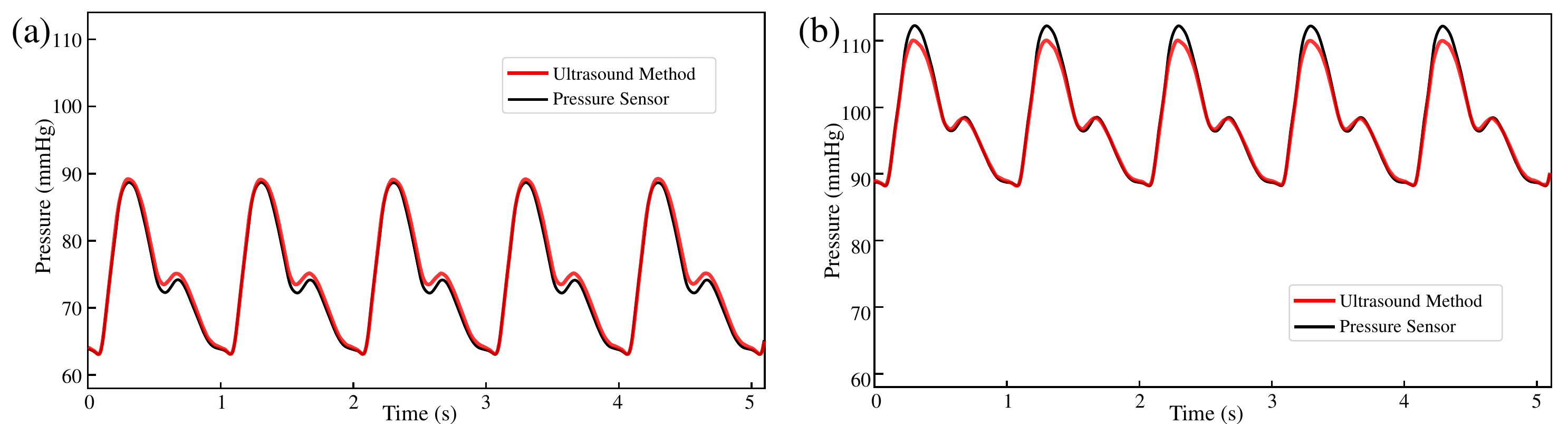}
\caption{Typical carotid artery blood pressure waveform measurement results. (a) DBP of 63mmHg.  (b) DBP of 88mmHg.
\label{fig_7}}
\end{figure*} 
\unskip

\begin{table}[ht]
\renewcommand{\arraystretch}{1.05} 
\caption{Results for the three experiment setups\label{tab:table1}}
\centering
\begin{tabular}{ c c c c c c }
\toprule
\makecell[c]{Trail \\ ID} & \makecell[c]{PP Error \\ (mmHg)} &\makecell[c]{RMSE \\ (mmHg)} & r & \makecell[c]{PWV$_U$ \\ (m/s)} & \makecell[c]{PWV$_R$ \\ (m/s)}\\
\midrule
1 & 2.83±0.04 & 1.50 & 0.995 & 7.88±0.62 & 8.02\\

2 & 0.53±0.05 & 0.91 & 0.998 & 8.32±0.59 & 8.03\\

3 & 2.25±0.06 & 0.90 & 0.998 & 7.95±0.57 & 8.13\\
\bottomrule
\end{tabular}
\end{table}

\subsection{Accuracy of in vivo PP and waveform measurements}
The description of the results from the arterial tonometry and ultrasound-based PP measurement experiment is presented in Fig. 8. The scatter plot and linear fit of PP values (Fig. 8(a)) show that the measured PP values (42.0 ± 7.7 mmHg) are strongly positively correlated (r=0.90, p \textless 0.001) with the reference PP values (44.5 ± 7.4 mmHg). This suggests that the measurement method is reliable. A linear regression line has been fitted to the data, which shows a significant positive relationship between the measured and reference values. In addition, Bland-Altman analysis has been performed to compare the measurements of PP obtained by the two methods, and the results are presented in Fig. 8(b). The mean difference between the two methods was found to be -2.58 mmHg with a standard deviation of 3.36 mmHg. The confidence interval was -9.15 mmHg and 4.00 mmHg. The Bland-Altman plot suggests no significant bias between the two methods. Only one measurement point falls outside the limits of agreement, indicating that the ultrasound system can provide reliable and consistent measurements of PP values in comparison to the arterial tonometry.

\begin{figure*}[!t]
\centering
\includegraphics[width=18.1cm]{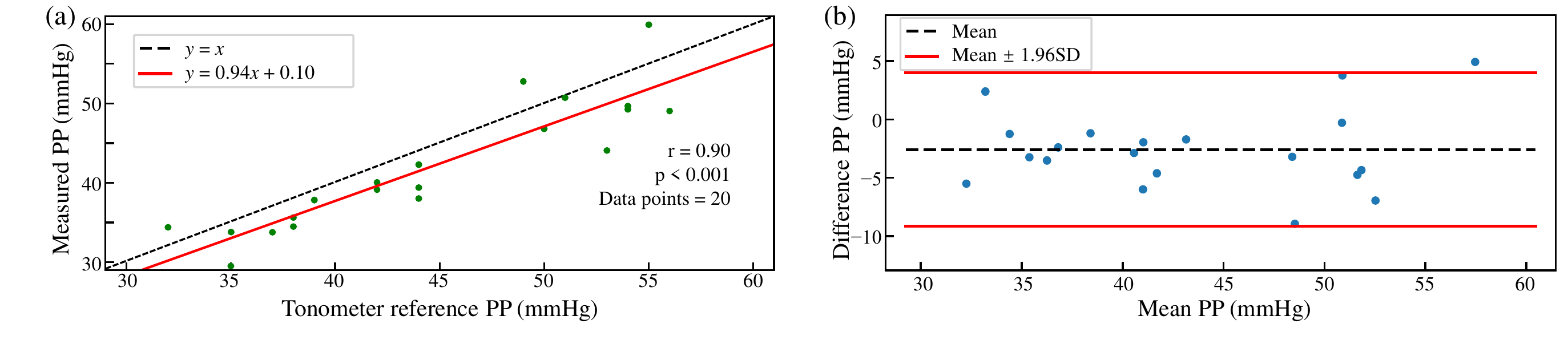}
\caption{(a) Scatter plot and linear regression analysis of measured PP and reference PP, the best fitting line is \textit{y} = 0.94 \textit{x} + 0.10. (b) Bland–Altman analysis comparing PP obtained by proposed technique and reference PP.
\label{fig_8}}
\end{figure*} 
\unskip

\begin{figure*}[!t]
\centering
\includegraphics[width=18.1cm]{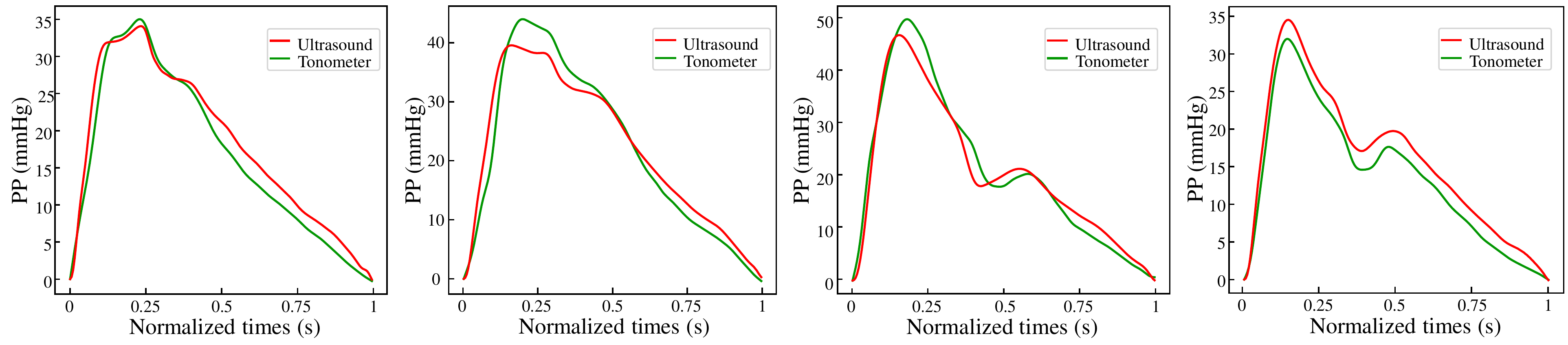}
\caption{Comparison of measured blood pressure
waveforms between the arterial tonometry and our ultrasound
measurement for four subjects.
\label{fig_9}}
\end{figure*} 
\unskip

\begin{figure}[!t]
\centering
\includegraphics[width=3in]{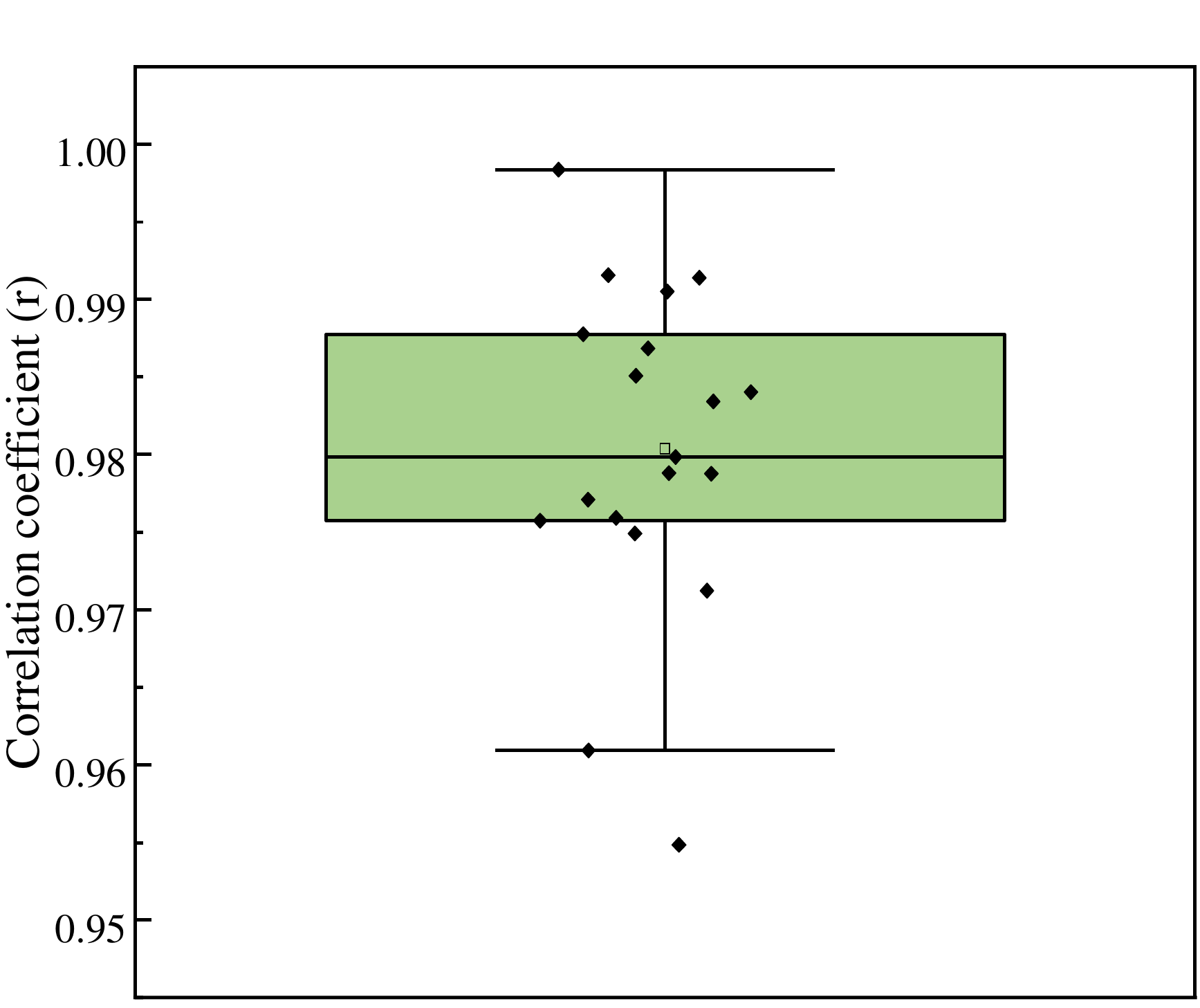}
\caption{Boxplot of correlation coefficient between measured waveform and reference waveform.
\label{fig_10}}
\end{figure} 
\unskip

Fig. 9 presents a comparison of the blood pressure waveforms measured by arterial tonometry and our ultrasound system for four subjects. The black line represents the average pressure waveform measured by the arterial tonometry, while the red line represents the ultrasound measurement in our study. The RMSE and waveform correlation (r) between the arterial tonometry and ultrasound measurement for all participants were 3.74 ± 1.61 mmHg and 0.978 ± 0.014, respectively.  Fig. 9 illustrates examples of PP waveforms obtained by these two methods in four human subjects. Additionally, a box plot (Fig. 10) was used to analyze the correlation coefficient values between the two pressure measurement methods, which included 20 values ranging from 0.931 to 0.998. The box plot indicated that 75\% of the values fell within the interquartile range (IQR) of 0.976 to 0.988, with a median value of 0.981. The lower whisker of the plot extended to 0.931 and the upper whisker extended to 0.998, indicating the presence of some outliers. Overall, the box plot suggested a high correlation between the two pressure measurement methods.

\subsection{Feasibility of tracing changing blood pressure in vivo}
In this subsection, the accuracy of the proposed ultrasound system for measuring variable blood pressure is described. To compare the accuracy of the ultrasound system with the volume clamp device Nexfin, the blood pressure value of Nexfin was transformed into carotid blood pressure. Based on the observation of constant DBP and MAP in the arterial tree\cite{b3, b4}, the DBP and MAP measured by Nexfin were used as the reference. The ultrasound blood pressure waveform of the first five beats was transformed into the carotid reference blood pressure waveform using the alternative calibration method\cite{b34, b41} and then to calculate its form factor (FF = (MAP - DBP) / PP), denoted as FF$_c$. The FF of the Nexfin finger blood pressure waveform of the first five beats was calculated, denoted as FF$_{fin}$. On subsequent cardiac cycles, the Nexfin pulse pressure PP$_{fin}$ is transformed to carotid reference pulse pressure PP$_{c-ref}$ by a linear transformation ( PP$_{c-ref}$ = PP$_{fin}$ × FF$_c$ / FF$_{fin}$)\cite{b41}, and then reference carotid SBP$_{c-ref}$ ( SBP$_{c-ref}$ = PP$_c$ + DBP ) is obtained, because DBP is constant in the arterial tree.

\begin{figure*}[!t]
\centering
\includegraphics[width=18.1cm]{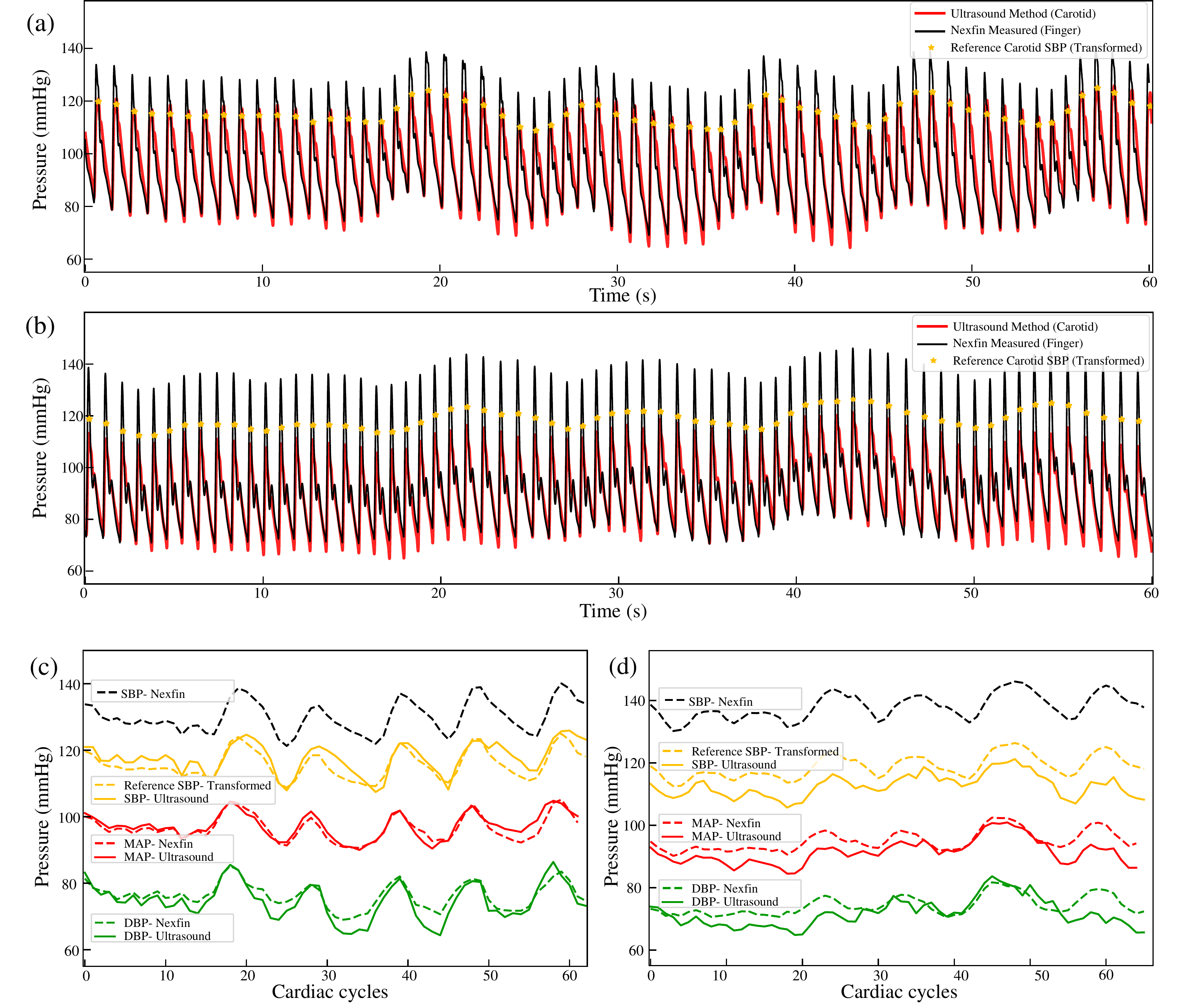}
\caption{ (a)(b) One-minute uninterrupted blood pressure waveforms of S4 and S6, respectively (c)(d) The beat-by-beat DBP, MAP and SBP changes of S4 and S6, respectively.
\label{fig_11}}
\end{figure*} 
\unskip
\begin{figure*}[!t]
\centering
\includegraphics[width=18.1cm]{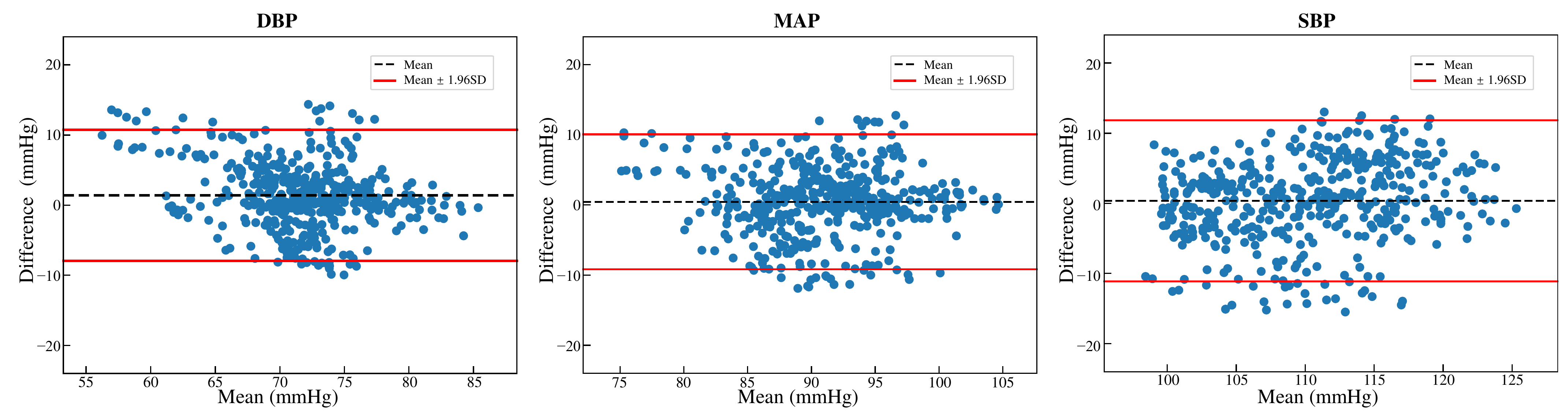}
\caption{Bland-Altman analysis comparing the measured blood pressure using ultrasound and the Nexfin device.
\label{fig_12}}
\end{figure*} 
\unskip

Fig. 11 shows the comparison between the measured and standard values of deep breathing blood pressure in two subjects. (S4 and S6). Fig. 11(a) and Fig. 11(b) show the one-minute uninterrupted blood pressure waveforms (the average DBP of the first five heartbeats measured by ultrasound was aligned to the value of Nexfin), and Figures Fig. 11(c) to Fig. 11(d) show the beat-by-beat DBP, MAP, and SBP changes compared to the reference values. During the normal breathing phase (0 - 15s), blood pressure remained relatively stable with no significant changes. While during the deep breathing phase (15 - 60s), blood pressure exhibited significant fluctuations in response to breathing. The continuous blood pressure measured by the ultrasound method in this work maintains the same trend of blood pressure changes as measured by Nexfin device. Table II demonstrates seven subjects' measurement errors (MAEs) of DBP, MAP, and SBP. On all data, the MAE of DBP, MAP, and SBP were 3.7±4.8 mmHg, 3.8±4.9 mmHg, and 4.6±5.9 mmHg, respectively. The accuracy of the proposed technique is shown in the Bland-Altman analysis (Fig. 12). The mean deviations of DBP, MAP and SBP were 1.39 mmHg, 0.44 mmHg and 0.36 mmHg, respectively, and the measured values within the consistency boundary accounted for 93.4 \%, 93.7 \% and 93.7 \%, respectively. The results show that the ultrasound measured values of blood pressure are in good agreement with the reference values.

\begin{table}[htpb]
\caption{Results for the three experiment setups\label{tab:table2}}
\centering
\renewcommand{\arraystretch}{1.00} 
\setlength{\tabcolsep}{17pt} 
\begin{tabular}{c ccc}
\toprule
\multirow{2}{*}{SubjectID} & \multicolumn{3}{c}{MAE (mmHg)}                                       \\ \cline{2-4} 
                           & \multicolumn{1}{c}{DBP}     & \multicolumn{1}{c}{MAP}     & SBP     \\ \hline
S1                         & \multicolumn{1}{c}{3.3±3.4} & \multicolumn{1}{c}{5.8±3.8} & 7.5±5.3 \\ 
S2                         & \multicolumn{1}{c}{6.7±4.1} & \multicolumn{1}{c}{5.1±3.5} & 5.0±4.2 \\ 
S3                         & \multicolumn{1}{c}{3.2±2.4} & \multicolumn{1}{c}{2.6±3.1} & 2.8±3.1 \\ 
S4                         & \multicolumn{1}{c}{2.2±2.1} & \multicolumn{1}{c}{1.2±1.5} & 2.3±2.0 \\ 
S5                         & \multicolumn{1}{c}{4.9±4.4} & \multicolumn{1}{c}{4.7±3.9} & 4.5±3.6 \\ 
S6                         & \multicolumn{1}{c}{3.4±3.3} & \multicolumn{1}{c}{3.6±2.4} & 6.2±2.3 \\ 
S7                         & \multicolumn{1}{c}{3.0±3.9} & \multicolumn{1}{c}{3.5±4.2} & 4.6±4.6 \\ 
Total                      & \multicolumn{1}{c}{3.7±4.8} & \multicolumn{1}{c}{3.8±4.9} & 4.6±5.9 \\ 
\bottomrule
\end{tabular}

\end{table}

\section{Discussion}
In this study, we proposed a three-element ultrasound system for real-time and noninvasive continuous measurement of beat-by-beat arterial PP and blood pressure waveforms without calibration. The consistency of DBP in the arterial tree and its ease of measurement by the brachial sphygmomanometer enables the system to obtain absolute blood pressure values and waveforms by inputting DBP values. Our proposed system's measurement capability was validated through well-established experiments, including in vitro cardiovascular phantom experiments and human experiments compared to arterial tonometer and volume clamp device. The results of all experiments demonstrate that our proposed system achieves continuous and accurate noninvasive measurement of beat-by-beat arterial PP and blood pressure waveforms. Beat-by-beat blood pressure waveform measurement is of clinical value for preventing and diagnosing CVDs. The proposed three-element system achieves calibration-free measurement of carotid PP and blood pressure waveforms and captures blood pressure changes continuously and accurately under deep breathing actions, providing a new means for diagnosing and preventing CVDs and extending the application scenario of blood pressure waveform monitoring.

In addition to PP and blood pressure waveform values, the proposed technique can obtain additional information about the cardiovascular system. For example, the local PWV measured by the present technology is an independent predictor of atherosclerosis and can reflect the degree of sclerosis at a specific location.
Combined with the measured arterial diameter, potential parameters of cardiovascular health can be calculated, such as pressure-strain elastic modulus (Ep = PP/($\Delta$D/Dd)\cite{b44} 
and local arterial compliance (AC = $\Delta$/PP). $\Delta$D is the difference between end-systolic and end-diastolic diameters, Dd is the end-diastolic diameter, and $\Delta$A is the difference between end-systolic and end-diastolic arterial cross-sectional area. These parameters are influenced by cardiovascular geometry and significantly affect cardiac load, with clinical implications for diagnosing and preventing CVDs. Although this work achieved satisfactory results, several limitations still need to be addressed to improve ease of use and accuracy, including mainly ultrasound measurements and BP estimation models.

Firstly, the proposed three-element system is an image-free technique that significantly improves frame rates and reduces computational effort. However, the lack of image guidance presents some operational difficulties. Although the software can automatically identify the arterial wall and calculate the SNR, the operator still needs to adjust the probe position based on the echo intensity and morphology, which requires proficiency. After approximately two weeks of use, operators typically require 30-60 seconds to obtain a reliable signal after palpating for the approximate carotid position. This time can be reduced to 5-20 seconds if the exact location of the carotid artery is marked using a B-Mode ultrasound system before testing. Complementary technologies are being developed to overcome these limitations and increase the system's usability. Due to the simplicity and scalability of the system, it can be easily embedded into conventional ultrasound imaging systems, similar to a multi-M-line ultrasound system. Emitting multiple M-Line ultrasounds between focused imaging improves ease of use with image guidance and increases the number of emitted array elements to improve the accuracy of local PWV. Since this technology is computationally small, only minimal additional computational resources are needed to complete the real-time or quasi-real-time evaluation of PP and blood pressure waveforms.

Another potential limitation of the proposed technology is the underlying physical assumptions in the blood pressure waveform estimation model, specifically the assumption that the artery is a cylindrical geometry and purely elastic tube. These two assumptions run through the Bramwell-Hill equation and the Moens-Korteweg equation, which are fundamental to ultrasound manometry theory assumptionscite\cite{b25, b30}.
While the assumption of an approximately axisymmetric circular tube applies to large arteries such as the carotid artery, it does not apply to arterial branches, aneurysms, and severe arterial plaques, limiting the technology's application. Additionally, the assumption of a purely elastic tube ignores the viscoelasticity term, which is negligible in healthy subjects but becomes apparent in some arterial sub-health populations, such as the elderly and atherosclerotic patients\cite{b46}.
Viscoelasticity causes the distension waveform to lag behind the pressure waveform and leads to nonlinear propagation of the PWV, complicating the relationship between pressure and diameter changes\cite{b26}. This may result in an underestimation of the overall pulse wave velocity if the PWV is measured at the foot of the distension waveform as the reference point\cite{b48}. Future work is needed to evaluate the accuracy of the proposed technology in broader application scenarios, including experiments on subjects with different health conditions and age groups, and to make corrections to the model.

Although our preliminary system was applied to the carotid artery, it has the potential to be extended to other superficial sites to measure PP and blood pressure waveforms, such as the brachial artery. It can also be considered for application on deep arteries, such as the abdominal aorta, as well as in combination with transesophageal echocardiography technology for measurements of the aorta. The wearability of ultrasound arrays is one of the future directions of ultrasound. The ability to conform to the skin for long-term measurements is achieved by flexibilities the arrays\cite{b28, b49} 
to adapt to complex skin surfaces. In previous work, our team developed a flexible ultrasound array with a 4 × 4 structure for local PWV measurements\cite{b49}. Given to the simplicity and easy scalability of the proposed technology, it requires only a few independent ultrasound elements to complete the measurements. It can be easily deployed on such flexible ultrasound arrays. Our future work includes incorporating this technique into previously developed flexible ultrasound array\cite{b49} to enable wearable PP and blood pressure waveforms measurements.

\section{Conclusion}
In this work, we developed a three-element image-free ultrasound system to enable the real-time and noninvasive continuous measurement of arterial PP and blood pressure waveform without calibration. It measures arterial distension directly through ultrasound penetration of tissue to acquire arterial diameter waveform and real-time local PWV with a low computational consumption local PWV evaluation method. It is a simple, easily scalable and can be easily extended to existing ultrasound systems and wearable ultrasound transducers. Its measurement capability was validated by in vitro experiments on a self-constructed cardiovascular phantom, and by in vivo experiments on human carotid arteries. In three groups of in vitro experiments, the error was less than 3 mmHg compared with the reference PP obtained by the invasive pressure sensor. In terms of waveform accuracy compared to the pressure sensor, the RMSE is within 2 mmHg, and the correlation coefficient is above 0.99. In-vivo PP and waveform accuracy comparison experiments with arterial tonometer were performed on 20 young, healthy subjects. Bland-Altman's analysis of estimates from the proposed method and arterial tonometer PP yielded an insignificant deviation of -2.58 mmHg with an error SD equal to 3.36 mmHg. Blood pressure waveforms compared with arterial tonometer demonstrated a high correlation (r = 0.978 ± 0.014). In comparative experiments involving continuous monitoring of deep breathing movements over a 60-second period in 7 young, healthy subjects, the MAE of DBP, MAP, and SBP was found to be within 5±8 mmHg (3.7±4.8 mmHg, 3.8±4.9 mmHg, and 4.6±5.9 mmHg respectively) compared to the volume clamp device, across all 442 cardiac cycles. Despite some limitations that need to be addressed for clinical usability, the results of the current study demonstrate the accuracy and reliability of the developed three-element ultrasound system for PP and blood pressure waveform measurements.

\end{document}